\documentclass[conference,10pt]{IEEEtran}
\IEEEoverridecommandlockouts
\usepackage{cite}
\usepackage{amsmath,amssymb,amsfonts}
\usepackage{graphicx}
\usepackage{textcomp}
\usepackage{amsmath}
\usepackage{booktabs}
\usepackage{stfloats}
\usepackage{setspace}
\usepackage{float}% hyperlinks
\usepackage{subcaption}
\usepackage[font=footnotesize,skip=0pt]{caption}
\usepackage[table]{xcolor}
\usepackage{booktabs}
\usepackage{adjustbox}
\usepackage{makecell}

\newcommand{\best}[1]{\cellcolor{green!18}\textbf{#1}}

\usepackage{multirow}
\usepackage{hyperref}       % hyperlinks
\usepackage{url}            % simple URL typesetting
\usepackage{booktabs}       % professional-quality tables
\usepackage{amsfonts}       % blackboard math symbols
\usepackage{nicefrac}       % compact symbols for 1/2, etc.
\usepackage{microtype}      % microtypography
\usepackage{xcolor}         % colors
\usepackage{algorithm}
\usepackage{algpseudocode}
\usepackage{amsmath}
\usepackage{tikz}

\usepackage{orcidlink}
\usepackage{comment}
\usepackage{amssymb}
\usepackage{braket}
\usepackage{xcolor}
\def\BibTeX{{\rm B\kern-.05em{\sc i\kern-.025em b}\kern-.08em
    T\kern-.1667em\lower.7ex\hbox{E}\kern-.125emX}}

\usepackage[compact]{titlesec}
\usepackage{stfloats}
\titlespacing\section{0pt}{0.2\baselineskip}{0.12\baselineskip}
\titlespacing\subsection{0pt}{0.15\baselineskip}{0.08\baselineskip}
\titlespacing\subsubsection{0pt}{0.1\baselineskip}{0.08\baselineskip}    
\begin{document}
\bstctlcite{IEEEexample:BSTcontrol}
\title{%Empirical Scaling Laws for Quantum Neural Networks in Classification Tasks
%Scaling Hybrid Quantum Neural Networks Beyond Accuracy: Depth, Qubits, and Quantum Metrics
Scaling Laws for Hybrid Quantum Neural Networks: Depth, Width, and Quantum-Centric Diagnostics
}

%\author{\IEEEauthorblockN{Anonymous Authors}}
%\begin{comment}
\author{\IEEEauthorblockN{Danil Vyskubov\textsuperscript{1}, Kirill Vyskubov\textsuperscript{1}, Nouhaila Innan\textsuperscript{2,3}, and Muhammad Shafique\textsuperscript{2,3}}

\IEEEauthorblockA{
\textsuperscript{1}University of Sharjah, Sharjah, UAE\\
\textsuperscript{2}eBRAIN Lab, Division of Engineering, New York University Abu Dhabi (NYUAD), Abu Dhabi, UAE\\
\textsuperscript{3}Center for Quantum and Topological Systems (CQTS), NYUAD Research Institute, NYUAD, Abu Dhabi, UAE\\
U22100183@sharjah.ac.ae, U22100221@sharjah.ac.ae, nouhaila.innan@nyu.edu, muhammad.shafique@nyu.edu\\
}}
%\end{comment}

\maketitle

\begin{abstract}
 Hybrid quantum neural networks are increasingly explored for classification, yet it remains unclear how their performance and quantum behavior scale with circuit depth and qubit count. We present a controlled scaling study of hybrid quantum-classical classifiers along two axes: (1) increasing the number of quantum layers $L$ at fixed qubits $Q$, and (2) increasing the number of qubits $Q$ at fixed depth $L$. Across multiple datasets, we evaluate predictive performance using Accuracy, PR-AUC, Precision, Recall, and F1, and track quantum-specific metrics (QCE, EEE, QGN) to characterize how quantum properties evolve under scaling. Our results summarize scaling trends, saturation regimes, and dataset-dependent sensitivity, and further analyze how quantum metrics relate to predictive performance. This study provides practical guidance for selecting $(Q,L)$ in hybrid QNN classifiers and establishes a consistent evaluation protocol for scaling analysis.
\end{abstract}

\begin{IEEEkeywords}
Quantum Machine Learning, Scaling Laws, Quantum Neural Networks, Classification 
\end{IEEEkeywords}

\section{Introduction}
Hybrid quantum neural networks (QNNs) combine classical neural modules with parameterized quantum circuits (PQCs) and are a widely used paradigm for quantum machine learning in the NISQ era \cite{Preskill_2018,innan2025next,choudhary2025hqnn,innan2025qnn,dave2025sentiqnf,innan2025circuithunt,10651123,innan2025quantum}. In such hybrid models, two architectural dimensions primarily determine the capacity of the quantum component: circuit width, given by the number of qubits $Q$, and circuit depth, given by the number of repeated variational layers $L$. Increasing $Q$ enlarges the accessible Hilbert space exponentially, whereas increasing $L$ increases the number of trainable transformations and entangling operations applied to the encoded data.

Scaling $Q$ and $L$, however, does not guarantee improved generalization. Deeper PQCs can exhibit optimization pathologies such as vanishing gradients (barren plateaus) \cite{McClean_2018}, while wider circuits increase parameter count and quantum complexity, which can lead to diminishing returns under a fixed training budget and in noisy settings. Consequently, principled QNN design requires understanding when architectural scaling improves predictive performance and when it primarily changes trainability or circuit-level behavior.

A key limitation in the current empirical literature is that scaling is often assessed using predictive metrics alone, or under protocols that simultaneously vary multiple factors alongside circuit size, such as preprocessing capacity, readout complexity, and training budget. This makes it difficult to attribute performance changes to the quantum architecture itself. Quantum-centric diagnostics such as expressibility and entanglement measures, along with gradient-based indicators of trainability, provide complementary visibility into how circuit families behave as $Q$ and $L$ grow, and help interpret saturation and instability beyond accuracy trends.

\begin{figure}[htpb]
    \centering
    \includegraphics[width=1\linewidth]{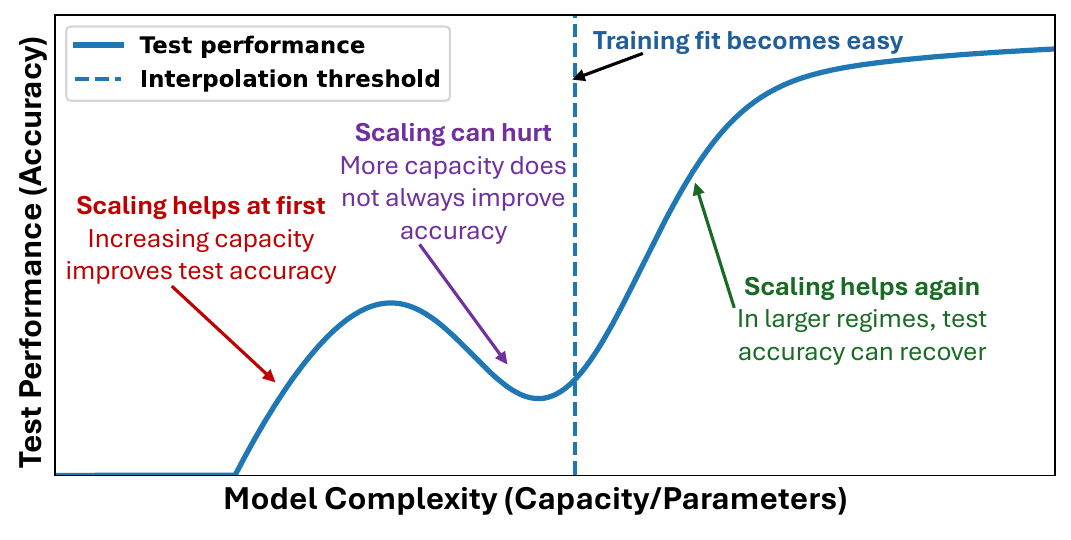}
    \caption{Schematic illustration of double descent in classical learning: test error may improve, degrade near the interpolation threshold, and improve again as capacity increases.}
    \label{fig:double_descent}
\end{figure}

In classical ML, test error can exhibit non-monotonic dependence on model capacity, including non-monotonic regimes often discussed under the double descent phenomenon \cite{Belkin_2019}. While hybrid QNNs differ from classical over-parameterized networks, the classical literature highlights an important methodological point: scaling trends are often structured but need not be monotonic (see Fig.~\ref{fig:double_descent}). This motivates a controlled, measurement-rich scaling analysis for hybrid QNN classifiers that varies one architectural axis at a time while keeping the remaining components fixed.

This work studies hybrid QNNs for supervised multi-class image classification under controlled architectural scaling of the quantum component. Each model consists of a classical preprocessing module that maps an input image to a $Q$-dimensional feature vector, a PQC with $Q$ qubits and $L$ variational layers using a matched entangling topology across experiments, and a lightweight classical readout head, trained end-to-end under a fixed optimizer and training budget per dataset. We address three questions:
\begin{itemize}
    \item \textit{How does predictive performance change as depth increases ($L\uparrow$) at fixed width ($Q$ fixed)?}
    \item \textit{How does predictive performance change as width increases ($Q\uparrow$) at fixed depth ($L$ fixed)?}
    \item \textit{How do quantum-centric diagnostics of expressibility, entanglement, and optimization dynamics evolve under each scaling regime, and how do they relate to predictive performance?}
\end{itemize}

Across three benchmarks, the study reveals dataset-dependent scaling regimes, including performance saturation and non-monotonic depth trends under fixed budgets, and shows that changes in width more consistently align with expressibility and entanglement growth, whereas changes in depth more strongly reflect optimization dynamics.

This paper makes the following contributions:
\begin{itemize}
    \item A controlled depth and width scaling study of hybrid QNN classifiers on three image benchmarks, varying one architectural axis at a time under fixed training budgets and matched circuit topology.
    \item A unified evaluation protocol that reports both predictive metrics and quantum-centric diagnostics to support interpretable scaling analysis.
    \item An empirical characterization of dataset-dependent scaling regimes, including saturation and non-monotonic depth behavior, contrasting the effects of increasing $L$ versus increasing $Q$.
    \item A performance--diagnostic tradeoff analysis based on rank correlation to quantify alignment between predictive gains and circuit diagnostics across scaling configurations.
\end{itemize}

The remainder of the paper is organized as follows: Sec. \ref{sec2} reviews related work; Sec. \ref{sec3} presents the problem setup and methodology; Sec. \ref{sec4} reports scaling results across datasets; Sec. \ref{sec5} concludes with key takeaways for scalable QNN design.

\section{Related Work and Background}\label{sec2}
\subsection{Hybrid Quantum Neural Networks}
\begin{figure}[htpb]
    \centering
    \includegraphics[width=1\linewidth]{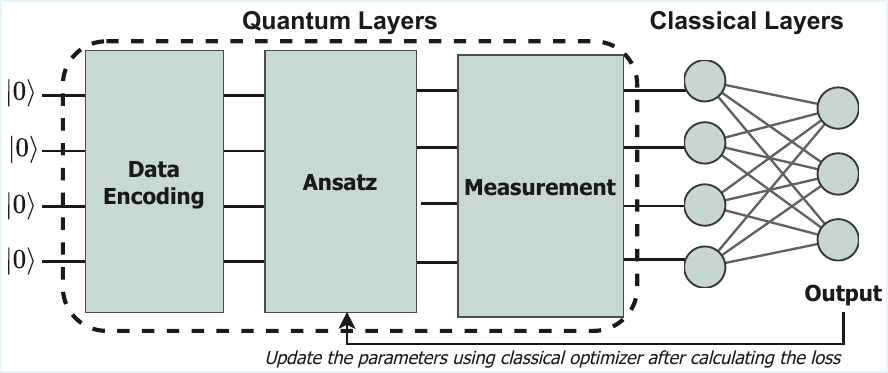}
    \caption{Generic hybrid QNN template. Input features are encoded into a $Q$-qubit PQC, measured to obtain classical statistics, and mapped to task outputs by a classical readout head. }
    \label{fig:hqnn_template}
\end{figure}

Hybrid QNNs are commonly constructed as end-to-end trainable pipelines in which a PQC is integrated within a classical learning system. A typical design maps input data to a low-dimensional set of features, applies a variational quantum transformation, and then uses a classical readout to produce task outputs \cite{Schuld_2014, Benedetti_2019}. Hybridization appears in two ways. First, classical layers are often placed \emph{before} the PQC to perform preprocessing and dimensionality reduction, since near-term circuits operate on a limited number of qubits and cannot directly ingest high-dimensional inputs. This stage is usually lightweight and aims to generate a $Q$-dimensional vector that can be embedded into the circuit via a chosen encoding scheme. Second, classical layers are commonly placed \emph{after} the PQC to map measurement outcomes to logits or regression targets (see Fig.~\ref{fig:hqnn_template}). This readout is also motivated by the fact that training is typically driven by classical optimization routines, where gradients are estimated from measurement statistics and propagated through the hybrid model.
Within this template, substantial architectural variation exists across the literature. Some models use only a post-quantum classical head, directly encoding raw inputs into the PQC, while others include both pre-quantum preprocessing and post-quantum readout. Encodings are frequently angle-based, and hardware-efficient ansatz designs based on trainable single-qubit rotations interleaved with fixed entangling patterns are widely adopted due to their compatibility with near-term devices \cite{Kandala_2017}. 

\subsection{Scaling, Expressivity, Trainability, and Quantum Metrics}
A central question in QML is how QNN behavior changes as circuits scale in width $Q$ (qubits) and depth $L$ (variational layers). Increasing $Q$ expands the Hilbert space $\dim(\mathcal{H}) = 2^{Q}$, while increasing $L$ increases the number of trainable transformations and entangling operations. Although both can increase capacity, their effects are not equivalent and often depend on the dataset and training regime.

Understanding scaling is challenging because expressivity and trainability can move in opposite directions. Larger circuits may represent richer quantum state families \cite{Sim_2019}, yet optimization can become harder. For broad classes of parameterized circuits and global cost functions, gradient statistics can deteriorate with size; in particular, the gradient variance may decay exponentially with qubit count, $\mathrm{Var}\!\left[\frac{\partial \mathcal{L}}{\partial \theta_i}\right]\in\mathcal{O}(2^{-Q})$, which is associated with barren plateaus and unstable training \cite{McClean_2018, Beer_2020}. Under fixed training budgets, these effects can lead to saturation, variability, or degradation in predictive performance despite increased capacity.

Because these mechanisms are difficult to disentangle from accuracy curves alone, recent work has emphasized the need for circuit-level diagnostics that quantify expressivity, entanglement structure, and optimization dynamics. Metrics such as quantum circuit expressibility \cite{Sim_2019}, effective entanglement entropy \cite{Meyer_2002}, and gradient-based diagnostics linked to trainability \cite{McClean_2018} provide measurable proxies for these latent properties and can explain why certain scaling choices help or fail. In this paper, we use such quantum-centric metrics to quantify how circuit behavior changes with $Q$ and $L$, and to relate these changes to predictive performance under controlled training conditions.

This perspective differs from much of the existing empirical literature, which commonly studies circuit size to identify strong task-specific settings, often tuning multiple components with circuit size and reporting predictive metrics as the primary outcome \cite{10.1007/978-3-031-85884-0_9}. Such comparisons are useful for application-level benchmarking, but they can obscure scaling mechanisms and make it difficult to attribute observed gains or failures to depth, width, or optimization effects. Our goal is instead to characterize scaling behavior itself: we vary one architectural axis at a time under matched circuit topology and fixed training budgets, and we jointly analyze predictive performance with quantum-centric metrics. This fills a gap by providing a controlled, diagnostic scaling study that connects depth and width trends to measurable changes in expressivity, entanglement, and gradient behavior.

\section{Methodology}\label{sec3}
\begin{figure*}[htpbt]
    \centering
    \includegraphics[width=1\linewidth]{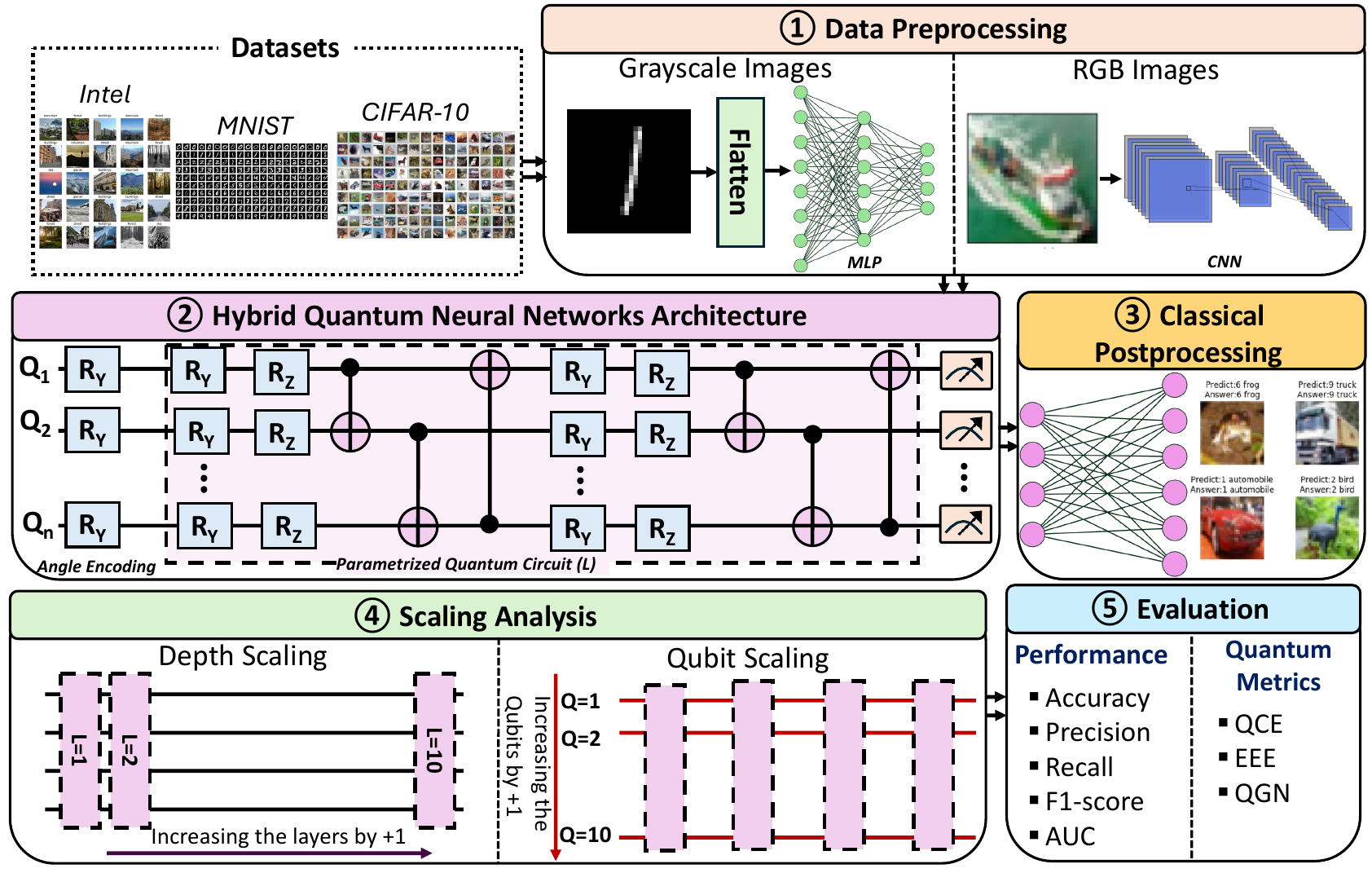}
    \caption{Overview of the controlled scaling protocol. Images are mapped to $Q$ features by a lightweight classical frontend, embedded into a $Q$-qubit PQC with $L$ layers, measured to obtain classical statistics, and decoded by a small classical head. Scaling is evaluated by varying one axis ($Q$ or $L$) while holding the others fixed.}
    \label{fig:methodology}
\end{figure*}

This section describes the end-to-end pipeline used to isolate depth and width scaling effects in hybrid QNN classifiers. Fig.~\ref{fig:methodology} summarizes the workflow: a lightweight classical frontend maps each image to a $Q$-dimensional feature vector, a $Q$-qubit PQC with $L$ variational layers transforms the encoded features and is measured to produce classical statistics, and a small classical head outputs class logits. We evaluate two controlled sweeps by varying one architectural axis at a time ($L$ at fixed $Q$, or $Q$ at fixed $L$), while keeping the remaining components and the training budget fixed per dataset.

\subsection{Data Preprocessing}
To embed high-dimensional images into a $Q$-qubit circuit, we use a lightweight classical frontend that produces a $Q$-dimensional real-valued feature vector suitable for angle embedding. The frontend capacity is kept modest so that performance differences across configurations primarily reflect the quantum component.

\subsubsection{Grayscale Data}
%For grayscale images, we use a multilayer perceptron (MLP). Each $28\times 28$ %image is flattened and passed through fully connected layers to produce a $Q$-%dimensional latent vector, which is used as the PQC input.%
For grayscale images, we use a simple three-layer multilayer perceptron (MLP). Each $28\times 28$ image is flattened into a 784-dimensional vector, passed through a hidden layer ($\mathbb{R}^{784} \rightarrow \mathbb{R}^{128}$ with ReLU activation), and projected to the $Q$-dimensional quantum input space ($\mathbb{R}^{128} \rightarrow \mathbb{R}^{Q}$). The intermediate dimension of 128 provides compressed features for the PQC input.

\subsubsection{RGB Data}

For RGB images, we use a compact six-layer convolutional neural network (CNN) to capture spatial structure and inter-channel correlations. Both CIFAR-10 (native $32\times 32$) and Intel Image Classification (originally $150\times 150$, resized to $32\times 32$) use identical preprocessing architectures to ensure fair comparison. The CNN consists of two convolutional blocks, each containing a $3\times 3$ convolution followed by ReLU activation and $2\times 2$ max-pooling. The first block maps from 3 to 32 channels, while the second maps from 32 to 64 channels, progressively reducing spatial dimensions from $32\times 32$ to $8\times 8$. The resulting $64 \times 8 \times 8 = 4096$-dimensional feature maps are flattened and compressed through two fully connected layers ($\mathbb{R}^{4096} \rightarrow \mathbb{R}^{128} \rightarrow \mathbb{R}^{Q}$) with ReLU nonlinearity.
In all cases, the resulting features are mapped to $[0,\pi]$ to ensure numerical stability and compatibility with angle-based data encoding.

\subsection{Hybrid Model Architecture}
We use a hybrid quantum--classical classifier in which a PQC acts as a trainable feature transformation layer. The dataset-dependent classical frontend and readout remain lightweight, while the PQC gate pattern and entangling topology are kept consistent across all experiments to support controlled scaling.

\subsubsection{Quantum Circuit Design}
The PQC operates on $Q$ qubits and contains $L$ repeated variational layers. Given an input $\mathbf{x}\in\mathbb{R}^{Q}$, the circuit applies $U(\boldsymbol{\theta}, \mathbf{x}) = U_{\text{var}}(\boldsymbol{\theta})\,U_{\text{enc}}(\mathbf{x}),$
where $U_{\text{enc}}$ encodes data and $U_{\text{var}}$ is the trainable ansatz.

\paragraph{Quantum Feature Encoding}
We use angle embedding with $Y$-axis rotations: $U_{\text{enc}}(\mathbf{x}) = \bigotimes_{i=1}^{Q} R_Y(x_i),$ with $x_i\in[0,\pi]$.

\paragraph{Variational Ansatz}
Each variational layer applies per-qubit trainable rotations followed by fixed entanglement: $R_Y(\theta_{\ell,i}^{(1)})\,R_Z(\theta_{\ell,i}^{(2)}),$ $i=1,\dots,Q,\ \ell=1,\dots,L,$
followed by a nearest-neighbor ring of CNOT gates (periodic boundary conditions for $Q>2$). The total number of quantum parameters scales as $|\boldsymbol{\theta}| = 2QL,$
yielding a hardware-efficient ansatz.

\paragraph{Measurement}
We measure Pauli-$Z$ expectation values on all qubits to obtain $\mathbf{z}=\left(\langle Z_1\rangle,\dots,\langle Z_Q\rangle\right)\in[-1,1]^Q.$

\subsection{Postprocessing Head}
The measured vector $\mathbf{z}$ is mapped to class logits by a small classical head. For grayscale datasets, we use a single linear layer $f(\mathbf{z})=W\mathbf{z}+\mathbf{b}$ to produce $C=10$ logits. For RGB datasets, we use a shallow MLP $\mathbb{R}^{Q}\rightarrow\mathbb{R}^{32}\rightarrow\mathbb{R}^{C}$ ($C=10$ for CIFAR-10, $C=6$ for Intel) with a ReLU nonlinearity. This keeps the readout low-capacity while allowing mild nonlinearity when required by the dataset.

\subsection{Controlled Scaling Protocol}
We evaluate two scaling regimes, varying one architectural axis at a time.

\subsubsection{Depth Scaling (Fixed $Q$, Vary $L$)}
To isolate depth effects, we fix the qubit count and vary the number of variational layers over $L\in\{2,\dots,10\}$ while keeping preprocessing, ansatz topology, optimizer type, and training budget fixed. Unless stated otherwise, depth sweeps use $Q=4$ for all datasets, selected as a stable operating point in pilot runs.

\subsubsection{Width Scaling (Fixed $L$, Vary $Q$)}
To isolate width effects, we fix the depth and vary the number of qubits over $Q\in\{2,\dots,10\}$. The classical frontend is resized only to output $Q$ features, while the encoding, ansatz pattern, and training protocol are unchanged. The fixed depth is selected for each dataset based on pilot runs that identify strong baselines, thereby enabling evaluation of width scaling around competitive operating points.

\subsection{Training and Optimization}
All hybrid models are trained end-to-end with a standard supervised objective. For each mini-batch, the classical frontend produces a $Q$-dimensional feature vector $\mathbf{x}\in[0,\pi]^Q$, which is encoded into the $Q$-qubit PQC, measured to obtain $\mathbf{z}\in[-1,1]^Q$, and mapped by the classical head to logits $\mathbf{o}\in\mathbb{R}^C$. Predicted probabilities are computed by softmax, and the cross-entropy loss is minimized: $\mathcal{L}(\Theta) = -\frac{1}{B}\sum_{i=1}^{B}\sum_{c=1}^{C}\mathbb{I}(y_i=c)\log p_{\Theta}(y=c\mid \mathbf{x}_i),$
where $B$ is the batch size and $\Theta$ denotes all trainable parameters (classical and quantum). Optimization is performed with Adam and $L_2$ weight decay using fixed hyperparameters per dataset. Gradients are backpropagated through the full hybrid pipeline; quantum parameter gradients are obtained through the differentiable quantum node in PennyLane and updated jointly with classical parameters in the same optimizer step. Within each dataset, the training budget (epochs and batch size) is kept constant across all $(Q,L)$ configurations to isolate architectural scaling effects. Model selection uses the checkpoint with the best validation performance, and final results are reported on the test set.

\subsection{Evaluation Metrics}
We report predictive metrics on the test set and quantum-centric diagnostics to interpret scaling behavior.

\subsubsection{Predictive Metrics}
We report accuracy, macro-precision, macro-recall, macro-F1, and PR-AUC. Accuracy is computed from argmax predictions: $\text{Accuracy}=\frac{1}{N}\sum_{i=1}^{N}\mathbb{I}(\hat{y}_i=y_i).$
For class $c$, precision and recall are computed from the confusion matrix,
$\text{Precision}_c=\frac{\mathrm{TP}_c}{\mathrm{TP}_c+\mathrm{FP}_c},$
$\text{Recall}_c=\frac{\mathrm{TP}_c}{\mathrm{TP}_c+\mathrm{FN}_c},$
and the F1-score is
$\text{F1}_c=2\cdot\frac{\text{Precision}_c\cdot\text{Recall}_c}{\text{Precision}_c+\text{Recall}_c}.$
%\vspace{0.3cm}
We report macro-averaged precision, recall, and F1 across classes. PR-AUC is computed from softmax probabilities using a one-vs-rest formulation with micro averaging.

\subsubsection{Quantum Metrics}
In addition to predictive performance, we report quantum-centric diagnostics that characterize circuit expressibility, entanglement structure, and optimization behavior \cite{ill}. Specifically, we report quantum circuit expressibility (QCE), effective entanglement entropy (EEE), and the quantum gradient norm (QGN).

\paragraph{Quantum Circuit Expressibility}
QCE quantifies how broadly a circuit family explores Hilbert space under random parameter draws. Following \cite{Sim_2019}, we compute QCE from the average pairwise fidelity between statevectors generated from $N$ random samples. QCE is evaluated using a statevector simulator on an auxiliary circuit that matches the gate pattern and entangling topology of the trained PQC: $\mathrm{QCE} = 1 - \frac{1}{N(N-1)} \sum_{i<j} \left|\langle\psi_i|\psi_j\rangle\right|^2.$

\paragraph{Effective Entanglement Entropy}
EEE measures entanglement between a subsystem and the remainder of the circuit. We compute the von Neumann entropy of the reduced density matrix $\rho_A$, as $S(\rho_A)=-\mathrm{Tr}(\rho_A\log\rho_A),$
where $A$ is chosen as the first $\lfloor Q/2\rfloor$ qubits and $\rho_A$ is obtained by tracing out the complement. As with QCE, EEE is computed from statevectors of an auxiliary circuit with the same structure as the trained PQC \cite{Meyer_2002}.

\paragraph{Quantum Gradient Norm}
QGN measures the magnitude of gradients associated with quantum parameters during training: $\mathrm{QGN}=\left\|\nabla_{\boldsymbol{\theta}_q}\mathcal{L}\right\|_2,$
where $\boldsymbol{\theta}_q$ denotes the quantum parameters. We extract these gradients from the hybrid training loop and report epoch-wise averages. Low values indicate weak update signals and can be consistent with vanishing-gradient regimes \cite{McClean_2018}.

\section{Results and Discussion}\label{sec4}

\subsection{Experimental Setup}
We evaluate hybrid QNN classifiers on MNIST \cite{726791}, CIFAR-10 \cite{krizhevsky2009learning}, and the Intel Image Classification dataset \cite{IntelImageClassification} under controlled architectural scaling. For each benchmark, the scaling studies are conducted around a dataset-specific operating point identified in pilot runs.  Concretely, the fixed axis in each sweep (the fixed $Q$ for the depth sweep and the fixed $L$ for the width sweep) is chosen to yield stable and competitive validation performance, ensuring that observed trends reflect scaling behavior rather than a suboptimal reference configuration. Table~\ref{tab:exp_setup} summarizes the dataset splits and hyperparameters used in all experiments.

\begin{table}[htpb]
\centering
\small
\caption{Experimental setup and hyperparameters for each dataset.}
\label{tab:exp_setup}
\begin{tabular}{lccc}
\hline
\textbf{Parameter} & \textbf{MNIST} & \textbf{CIFAR-10} & \textbf{Intel} \\
\hline
Fixed $L$ (qubit vary) & 6 & 2 & 2 \\
Fixed $Q$ (depth vary) & 4 & 4 & 4 \\
Classical Layers & 3 & 6 & 6 \\
Batch size & 64 & 128 & 64 \\
Epochs & 10 & 15 & 10 \\
Learning rate & $1\times10^{-2}$ & $1\times10^{-3}$ & $1\times10^{-3}$ \\
\hline
Training samples & 8{,}000 & 20{,}000 & 11{,}228 \\
Validation samples & 2{,}000 & 5{,}000 & 2{,}806 \\
Test samples & 5{,}000 & 10{,}000 & 3{,}000 \\
\hline
\end{tabular}
\end{table}

\subsection{Depth Scaling Analysis}
\begin{figure}[htpb]
    \centering
    \includegraphics[width=1.0\linewidth]{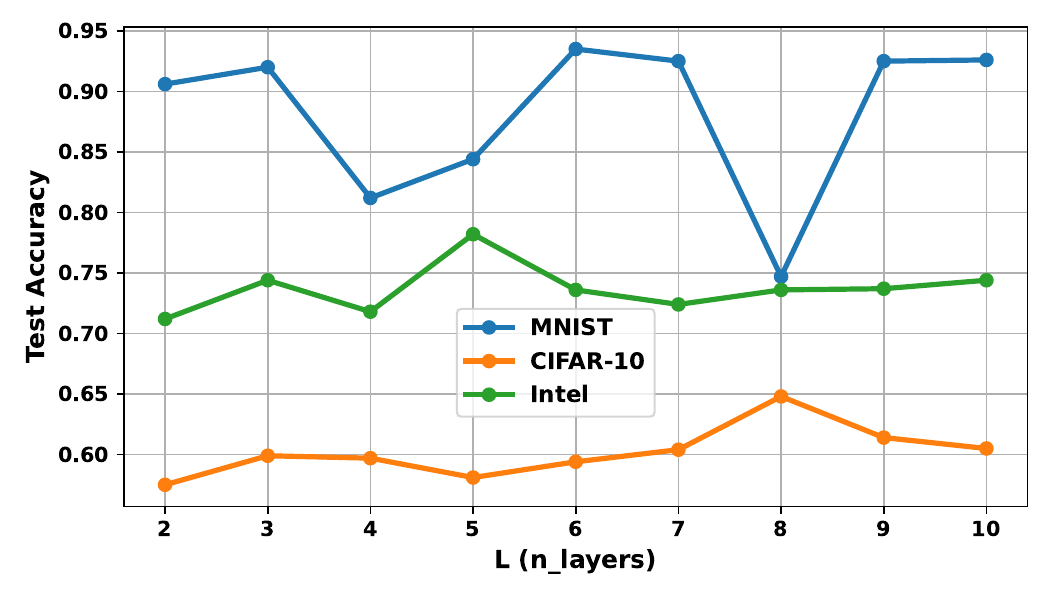}
    \caption{Depth scaling at fixed $Q=4$: test accuracy vs.\ layers $L$.}
    \label{fig:testacc_vs_L}
\end{figure}

\begin{figure}[htpbt]
    \centering
    \includegraphics[width=1.0\linewidth]{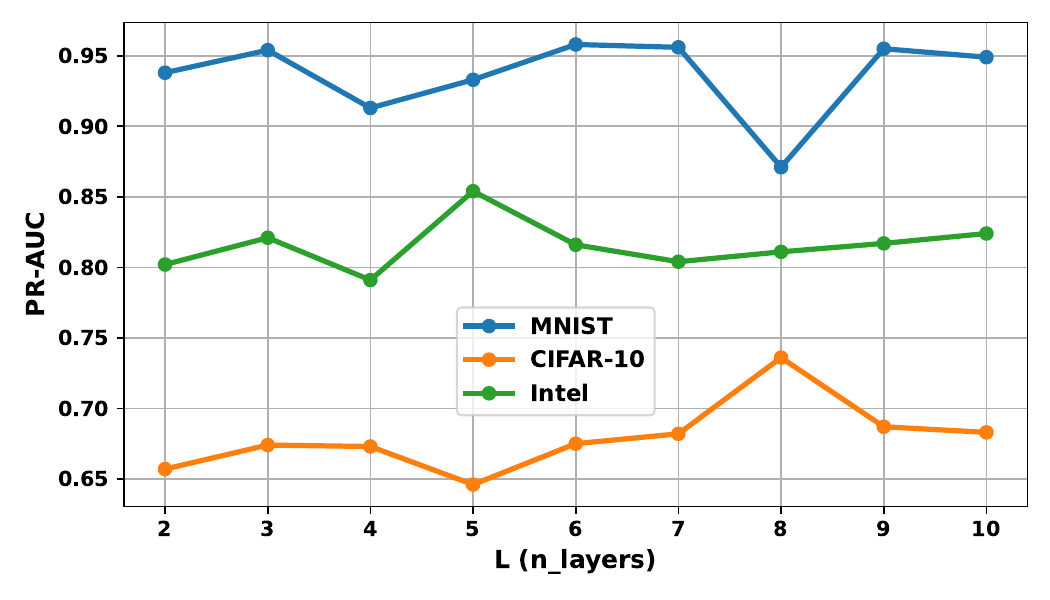}
    \caption{Depth scaling at fixed $Q=4$: PR-AUC vs.\ layers $L$.}
    \label{fig:pr-auc_vs_L}
\end{figure}

We first isolate the effect of circuit depth by fixing $Q=4$ and varying $L\in\{2,\dots,10\}$. Fig.~\ref{fig:testacc_vs_L} and Fig.~\ref{fig:pr-auc_vs_L} show that depth scaling is not uniformly beneficial: performance improves from shallow to moderate depths, but the trend becomes irregular once additional layers are introduced. This pattern is consistent across datasets, although the depth at which the curve flattens or oscillates differs by benchmark. Notably, CIFAR-10 benefits from additional depth up to a moderate range (peaking around $L{=}8$), whereas MNIST exhibits sharper swings with pronounced drops at intermediate depths (e.g., $L{=}4$ and $L{=}8$). Intel shows smaller-amplitude fluctuations, with its best point reached earlier (around $L{=}5$), followed by a largely saturated regime. Overall, the results indicate that adding layers can help initially, but deeper PQCs can become sensitive to optimization and do not yield predictable generalization gains under a fixed training budget.

\begin{figure*}[htpb]
    \centering
    \includegraphics[width=1.0\linewidth]{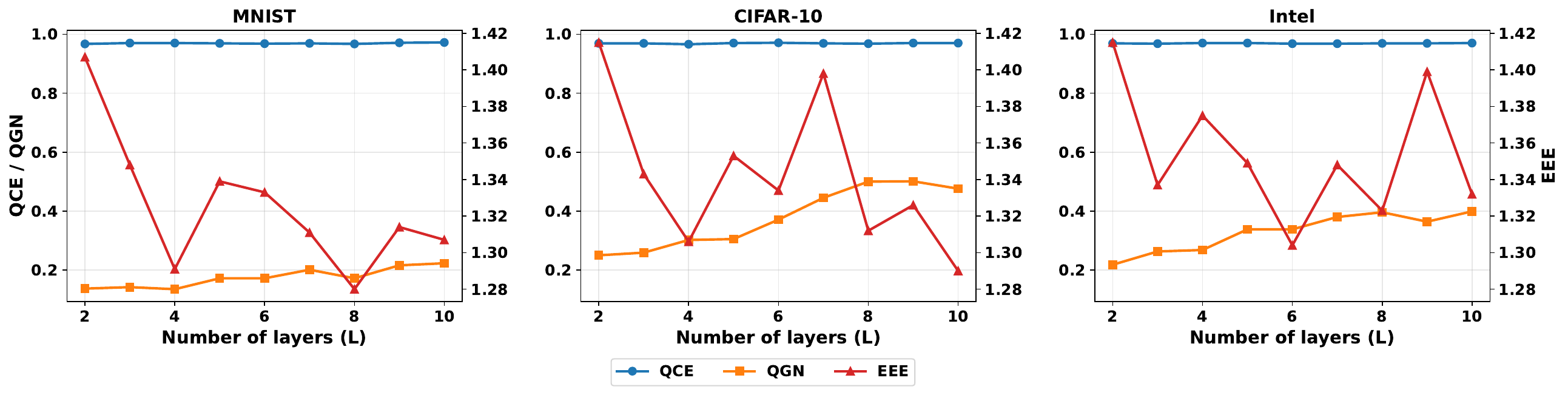}
    \caption{Depth scaling at fixed $Q=4$: QCE, EEE, and QGN vs.\ layers $L$.}
    \label{fig:qmetrics_layer_sweep}
\end{figure*}

Quantum diagnostics clarify why deeper circuits do not consistently help. As shown in Fig.~\ref{fig:qmetrics_layer_sweep}, QCE is already high at small $L$ and changes only marginally as depth grows, indicating that additional layers do not substantially broaden the explored state family at this width. Similarly, EEE rises quickly at shallow depth and then remains in a narrow, near-saturated band, suggesting that entanglement is established early and that extra layers primarily rearrange existing correlations rather than increasing entanglement capacity. In contrast, QGN becomes increasingly variable with $L$, reflecting greater sensitivity of the optimization dynamics as the circuit deepens. This combination is first early saturation of expressibility and entanglement proxies and second increasing variability in gradient magnitude, which is aligned with the non-monotonic accuracy/PR-AUC curves: once the circuit reaches a regime where representational gains are limited, deeper layers mainly affect training stability, leading to irregular performance. Table~\ref{tab:layer_sweep_all_side_compact} summarizes the full metrics and corroborates that the best-performing depths are dataset-specific and typically occur before the deepest configurations.

%%%%%%%%%%%%%%%%%%%%%%%%
\subsection{Qubit Scaling Analysis}
\begin{figure}[htpb]
    \centering
    \includegraphics[width=1.0\linewidth]{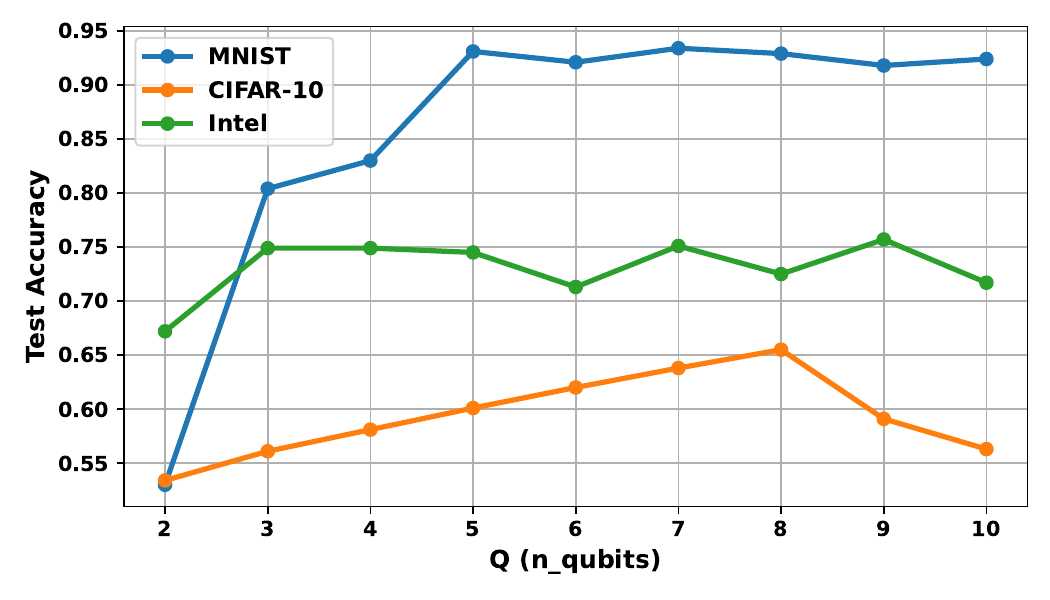}
    \caption{Width scaling at fixed $L$: test accuracy vs.\ qubits $Q$.}
    \label{fig:testacc_vs_Q}
\end{figure}

\begin{figure}[htpbt]
    \centering
    \includegraphics[width=1.0\linewidth]{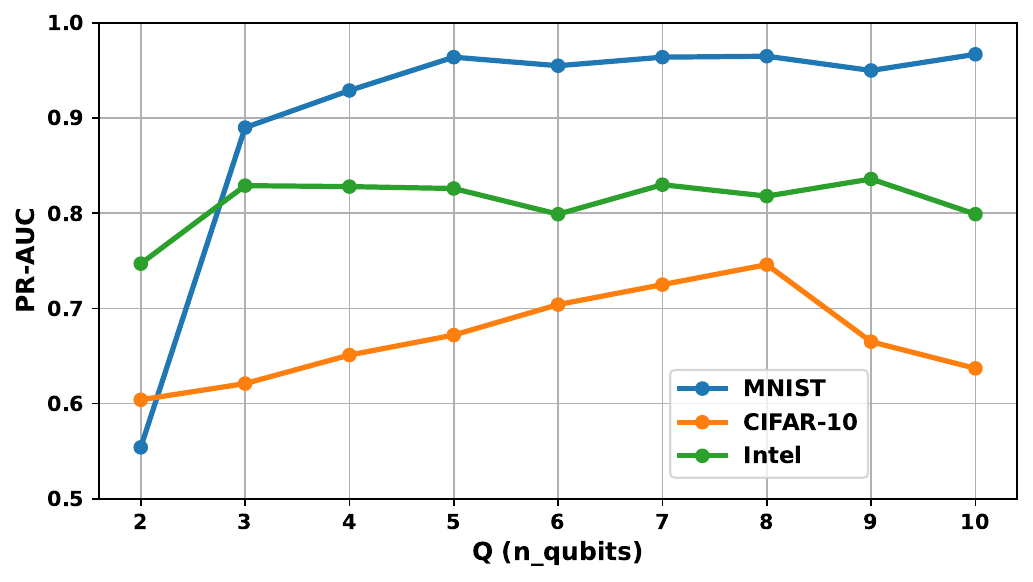}
    \caption{Width scaling at fixed $L$: PR-AUC vs.\ qubits $Q$.}
    \label{fig:pr-auc_vs_Q}
\end{figure}

We next isolate width effects by varying the number of qubits $Q\in\{2,\dots,10\}$ while keeping the circuit depth fixed per dataset. Fig.~\ref{fig:testacc_vs_Q} and Fig.~\ref{fig:pr-auc_vs_Q} show a markedly smoother scaling profile than the depth sweep: performance rises rapidly as the model moves from very small circuits to moderate width, then enters a saturation regime where additional qubits yield marginal gains and may occasionally reduce performance. This pattern is consistent across benchmarks but differs in where the plateau begins. For MNIST, accuracy increases sharply from $Q=2$ to $Q\approx 5$ and then remains near its peak, indicating that moderate width is sufficient to capture the relatively low-complexity decision structure. CIFAR-10 shows a more gradual improvement that continues up to intermediate-to-large widths, with the best point at $Q=8$ before degrading at $Q\ge 9$, suggesting that additional capacity helps until the fixed budget can no longer reliably optimize the larger hypothesis class. Intel exhibits the weakest monotonicity: performance improves early, then oscillates around a narrow band (peaking at $Q=9$), reflecting an intermediate regime where added capacity is beneficial but sensitive to training variability.

\begin{figure*}[htpb]
    \centering
    \includegraphics[width=1.0\linewidth]{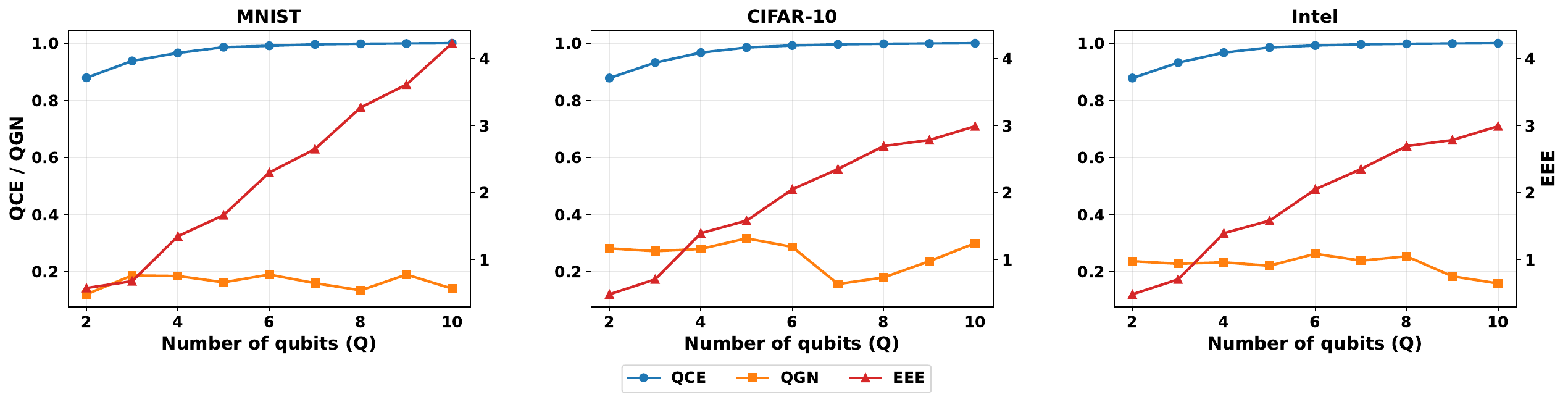}
    \caption{Width scaling at fixed $L$: QCE, EEE, and QGN vs.\ qubits $Q$.}
    \label{fig:qmetrics_qubit_sweep}
\end{figure*}

The quantum diagnostics closely track these trends and explain why width scaling is typically more predictable. As shown in Fig.~\ref{fig:qmetrics_qubit_sweep}, both QCE and EEE increase steadily with $Q$, indicating systematic growth in circuit expressibility and entanglement capacity as the Hilbert space expands. This contrasts with the depth sweep, where both metrics saturated quickly, and suggests that additional qubits more directly enlarge the set of accessible representations. Importantly, QGN remains within a comparable range across most widths, rather than becoming progressively more erratic, which is consistent with width adding capacity without uniformly weakening the training signal. The degradations observed at the largest widths for CIFAR-10 (and the mild oscillations on Intel) therefore align with a practical regime change: although expressibility and entanglement continue to grow, the fixed optimization budget and dataset complexity can make the enlarged parameter space harder to exploit reliably, leading to saturation or small regressions.  Table~\ref{tab:qubit_sweep_all_side_compact} summarizes the full set of predictive and quantum metrics, corroborating that the optimal width is dataset-specific and typically occurs at intermediate $Q$ before saturation or mild degradation at the largest qubit counts.

\begin{table*}[htpbt]
\centering
\small
\caption{Scaling results with compact formatting. Green highlights the best accuracy per dataset within each sweep.}
\label{tab:scaling_compact_colored}

\begin{subtable}[t]{0.49\linewidth}
\centering
\caption{Layer scaling (fixed $Q=4$)}
\label{tab:layer_sweep_all_side_compact}
\begin{adjustbox}{max width=\linewidth}
\begin{tabular}{l c c c c c c c c}
\toprule
\textbf{Data} & $L$ & Acc & Prec & Rec & F1 & QCE & EEE & QGN \\
\midrule
\rowcolor{gray!8}\multicolumn{9}{l}{\textbf{MNIST}}\\
 & 2  & 0.906 & 0.910 & 0.905 & 0.905 & 0.967 & 1.407 & 0.137 \\
 & 3  & 0.920 & 0.923 & 0.919 & 0.920 & 0.970 & 1.348 & 0.142 \\
 & 4  & 0.812 & 0.811 & 0.808 & 0.802 & 0.970 & 1.291 & 0.135 \\
 & 5  & 0.844 & 0.863 & 0.845 & 0.813 & 0.969 & 1.339 & 0.172 \\
 & 6  & \best{0.935} & 0.936 & 0.934 & 0.935 & 0.968 & 1.333 & 0.172 \\
 & 7  & 0.925 & 0.927 & 0.925 & 0.925 & 0.969 & 1.311 & 0.201 \\
 & 8  & 0.747 & 0.700 & 0.750 & 0.705 & 0.967 & 1.280 & 0.172 \\
 & 9  & 0.925 & 0.925 & 0.924 & 0.924 & 0.971 & 1.314 & 0.216 \\
 & 10 & 0.926 & 0.926 & 0.925 & 0.925 & 0.972 & 1.307 & 0.223 \\
\midrule
\rowcolor{gray!8}\multicolumn{9}{l}{\textbf{CIFAR-10}}\\
 & 2  & 0.575 & 0.571 & 0.575 & 0.567 & 0.969 & 1.415 & 0.250 \\
 & 3  & 0.599 & 0.599 & 0.599 & 0.592 & 0.969 & 1.343 & 0.259 \\
 & 4  & 0.597 & 0.598 & 0.597 & 0.584 & 0.966 & 1.306 & 0.302 \\
 & 5  & 0.581 & 0.563 & 0.581 & 0.560 & 0.970 & 1.353 & 0.305 \\
 & 6  & 0.594 & 0.592 & 0.594 & 0.587 & 0.971 & 1.334 & 0.371 \\
 & 7  & 0.604 & 0.587 & 0.604 & 0.590 & 0.969 & 1.398 & 0.445 \\
 & 8  & \best{0.648} & 0.647 & 0.648 & 0.643 & 0.968 & 1.312 & 0.500 \\
 & 9  & 0.614 & 0.612 & 0.613 & 0.603 & 0.970 & 1.326 & 0.501 \\
 & 10 & 0.605 & 0.614 & 0.605 & 0.602 & 0.970 & 1.290 & 0.476 \\
\midrule
\rowcolor{gray!8}\multicolumn{9}{l}{\textbf{Intel}}\\
 & 2  & 0.712 & 0.712 & 0.713 & 0.709 & 0.969 & 1.415 & 0.218 \\
 & 3  & 0.744 & 0.755 & 0.744 & 0.747 & 0.968 & 1.337 & 0.263 \\
 & 4  & 0.718 & 0.741 & 0.721 & 0.723 & 0.970 & 1.375 & 0.268 \\
 & 5  & \best{0.782} & 0.784 & 0.783 & 0.783 & 0.970 & 1.349 & 0.338 \\
 & 6  & 0.736 & 0.741 & 0.737 & 0.737 & 0.968 & 1.304 & 0.338 \\
 & 7  & 0.724 & 0.737 & 0.726 & 0.726 & 0.968 & 1.348 & 0.380 \\
 & 8  & 0.736 & 0.748 & 0.735 & 0.738 & 0.969 & 1.323 & 0.396 \\
 & 9  & 0.737 & 0.741 & 0.739 & 0.738 & 0.969 & 1.399 & 0.364 \\
 & 10 & 0.744 & 0.761 & 0.746 & 0.745 & 0.970 & 1.332 & 0.399 \\
\bottomrule
\end{tabular}
\end{adjustbox}
\end{subtable}\hspace{0.1pt}
\begin{subtable}[t]{0.49\linewidth}
\centering
\caption{Qubit scaling (vary $Q$, with $L$ as reported)}
\label{tab:qubit_sweep_all_side_compact}
\begin{adjustbox}{max width=\linewidth}
\begin{tabular}{l c c c c c c c c c}
\toprule
\textbf{Data} & $Q$ & $L$ & Acc & Prec & Rec & F1 & QCE & EEE & QGN \\
\midrule
\rowcolor{gray!8}\multicolumn{10}{l}{\textbf{MNIST}}\\
 & 2  & 6 & 0.530 & 0.544 & 0.521 & 0.421 & 0.879 & 0.577 & 0.121 \\
 & 3  & 6 & 0.804 & 0.821 & 0.804 & 0.797 & 0.938 & 0.679 & 0.187 \\
 & 4  & 6 & 0.830 & 0.787 & 0.831 & 0.800 & 0.966 & 1.350 & 0.185 \\
 & 5  & 6 & 0.931 & 0.931 & 0.931 & 0.931 & 0.986 & 1.665 & 0.163 \\
 & 6  & 6 & 0.921 & 0.922 & 0.920 & 0.921 & 0.991 & 2.301 & 0.190 \\
 & 7  & 6 & \best{0.934} & 0.933 & 0.933 & 0.933 & 0.996 & 2.652 & 0.160 \\
 & 8  & 6 & 0.929 & 0.929 & 0.928 & 0.928 & 0.998 & 3.273 & 0.135 \\
 & 9  & 6 & 0.918 & 0.920 & 0.918 & 0.918 & 0.999 & 3.615 & 0.190 \\
 & 10 & 6 & 0.924 & 0.925 & 0.923 & 0.924 & 1.000 & 4.232 & 0.141 \\
\midrule
\rowcolor{gray!8}\multicolumn{10}{l}{\textbf{CIFAR-10}}\\
 & 2  & 2 & 0.534 & 0.524 & 0.534 & 0.526 & 0.878 & 0.483 & 0.282 \\
 & 3  & 2 & 0.561 & 0.568 & 0.561 & 0.553 & 0.932 & 0.706 & 0.272 \\
 & 4  & 2 & 0.581 & 0.595 & 0.581 & 0.579 & 0.967 & 1.395 & 0.280 \\
 & 5  & 2 & 0.601 & 0.597 & 0.601 & 0.593 & 0.985 & 1.583 & 0.317 \\
 & 6  & 2 & 0.620 & 0.610 & 0.620 & 0.611 & 0.992 & 2.050 & 0.287 \\
 & 7  & 2 & 0.638 & 0.637 & 0.638 & 0.632 & 0.996 & 2.353 & 0.157 \\
 & 8  & 2 & \best{0.655} & 0.647 & 0.655 & 0.645 & 0.998 & 2.679 & 0.180 \\
 & 9  & 2 & 0.591 & 0.574 & 0.591 & 0.577 & 0.999 & 2.786 & 0.237 \\
 & 10 & 2 & 0.563 & 0.551 & 0.563 & 0.547 & 1.000 & 2.993 & 0.300 \\
\midrule
\rowcolor{gray!8}\multicolumn{10}{l}{\textbf{Intel}}\\
 & 2  & 2 & 0.672 & 0.661 & 0.674 & 0.659 & 0.878 & 0.483 & 0.237 \\
 & 3  & 2 & 0.749 & 0.764 & 0.749 & 0.751 & 0.932 & 0.706 & 0.228 \\
 & 4  & 2 & 0.749 & 0.752 & 0.751 & 0.749 & 0.967 & 1.395 & 0.233 \\
 & 5  & 2 & 0.745 & 0.755 & 0.746 & 0.746 & 0.985 & 1.583 & 0.221 \\
 & 6  & 2 & 0.713 & 0.732 & 0.713 & 0.719 & 0.992 & 2.050 & 0.263 \\
 & 7  & 2 & 0.751 & 0.754 & 0.752 & 0.753 & 0.996 & 2.353 & 0.239 \\
 & 8  & 2 & 0.725 & 0.750 & 0.730 & 0.727 & 0.998 & 2.679 & 0.254 \\
 & 9  & 2 & \best{0.757} & 0.762 & 0.757 & 0.757 & 0.999 & 2.786 & 0.184 \\
 & 10 & 2 & 0.717 & 0.746 & 0.722 & 0.723 & 1.000 & 2.993 & 0.159 \\
\bottomrule
\end{tabular}
\end{adjustbox}
\end{subtable}

\end{table*}
\subsection{Performance--Diagnostics Alignment}

While the previous subsections analyzed predictive performance and quantum diagnostics separately, we now examine whether improvements in predictive performance are systematically aligned with changes in intrinsic quantum properties. The goal of this analysis is to determine which diagnostics meaningfully track performance gains under architectural scaling, and whether this alignment differs between width- and depth-driven regimes.

\begin{figure*}[htpbt]
    \centering
    \includegraphics[width=1.0\linewidth]{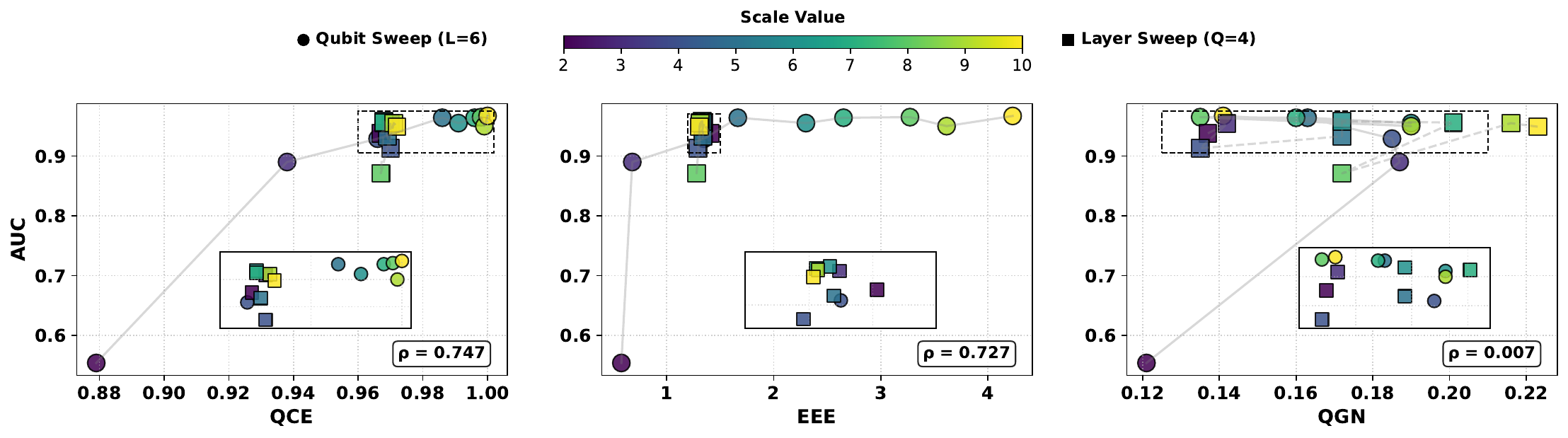}
    \caption{Spearman rank correlations between predictive metrics and quantum diagnostics across scaling configurations.}
    \label{fig:correlation_plot}
\end{figure*}

To this end, we compute Spearman rank correlations between AUC and each diagnostic across all scaling configurations, $\rho(P,M)=\mathrm{corr}\!\left(\mathrm{rank}(P),\mathrm{rank}(M)\right),$
and summarize the resulting relationships in Fig.~\ref{fig:correlation_plot}.
A clear ordering emerges for capacity-related quantities. As circuit capacity increases through width scaling, configurations with higher expressibility consistently achieve higher AUC, yielding a strong positive rank correlation. This reflects a structured regime in which expanding the accessible state space produces a reliable improvement in model ranking, rather than isolated performance spikes. In contrast, depth scaling contributes little to this ordering: performance fluctuations induced by additional layers occur without commensurate changes in expressibility, weakening the overall alignment.

A similar pattern holds for entanglement. Higher-performing configurations tend to occupy more entangled regimes, again driven primarily by width scaling. Once entanglement saturates under depth increases, further variation in performance is no longer explained by this quantity, indicating that additional layers do not systematically expand the effective representational structure explored by the circuit.

The optimization-related diagnostic behaves fundamentally differently. Its correlation with AUC is close to zero, indicating that gradient magnitude does not impose a consistent ranking over configurations. Instead, it reflects local training sensitivity: deeper circuits may exhibit larger or more variable gradients without achieving better generalization. This decoupling confirms that optimization difficulty alone does not explain which architectures perform best under a fixed budget.

The correlation structure reveals a separation of roles: width scaling drives performance gains that are coherently tracked by capacity-related diagnostics, while depth scaling primarily perturbs optimization dynamics without establishing a stable performance ordering. This explains why width produces smoother and more reliable gains across datasets, whereas depth leads to non-monotonic and dataset-sensitive behavior.

\subsection{Cross-Dataset Scaling Sensitivity}
Scaling sensitivity differs across datasets because circuit capacity interacts with image modality, class structure, and data availability in different ways. For MNIST (grayscale, simpler patterns, clearer class boundaries), strong performance is reached with moderate capacity, so further increases in depth mainly affect training behavior: QCE and EEE are already near a saturated regime, while QGN becomes more variable as $L$ grows, which matches the observed non-monotonic accuracy trends under a fixed budget. For CIFAR-10 and Intel (RGB, richer textures/backgrounds, higher intra-class variation), performance remains more representation-limited for longer. Increasing $Q$ produces systematic growth in QCE and EEE, and performance improvements track these increases until saturation, while QGN stays comparatively stable across most widths. Dataset size also modulates where saturation occurs: larger effective training sets typically sustain the benefits of added qubits longer, whereas smaller sets reach diminishing returns earlier and show stronger sensitivity to depth-induced QGN variability.

\subsection{Practical Takeaways}
\begin{itemize}
    \item \textbf{Favor width scaling for more predictable gains under fixed budgets.} Increasing $Q$ typically yields smoother improvements and is more consistently accompanied by increases in QCE/EEE than increasing $L$.
    \item \textbf{Depth has dataset-specific optima and can reduce stability.} After a moderate $L$, gains often saturate and may become non-monotonic, consistent with increased variability in QGN.
    \item \textbf{Use diagnostics to detect diminishing returns.} When QCE/EEE plateau and QGN becomes more variable, additional scaling is unlikely to provide reliable improvements.
    \item \textbf{Select scaling based on dataset characteristics.} Simpler datasets tend to saturate earlier, whereas more complex datasets benefit from additional qubits before reaching a plateau.
\end{itemize}

\section{Conclusion}\label{sec5}
This work studied scaling behavior in hybrid quantum neural networks for image classification through controlled sweeps over circuit depth $L$ and width $Q$ across multiple datasets, while holding the remaining components and training budgets fixed within each benchmark. The results highlight a clear separation between the two scaling axes. Increasing $Q$ generally yields more reliable performance gains and these gains tend to co-evolve with increases in QCE and EEE, indicating that width primarily drives measurable capacity growth. In contrast, increasing $L$ provides limited benefits beyond dataset-specific ranges and more frequently introduces instability, reflected by higher variability in QGN and non-monotonic predictive trends under fixed budgets. Across datasets, simpler settings reach strong performance with moderate capacity and show earlier saturation, whereas more complex settings benefit longer from width expansion before diminishing returns emerge. Overall, the findings suggest that effective scaling of hybrid QNNs requires balancing capacity-related growth (tracked by QCE/EEE) with optimization stability (tracked by QGN) in a dataset-aware manner, and motivate the use of quantum-centric diagnostics alongside predictive metrics when selecting scalable architectures.

\section*{Acknowledgment}
This work was supported in part by the NYUAD Center for Quantum and Topological Systems (CQTS), funded by Tamkeen under the NYUAD Research Institute grant CG008.

\begin{spacing}{0.92}
\bibliographystyle{IEEEtran}
\bibliography{refs}
\end{spacing}

\end{document}